# Coarse-grained structure of a physical (strange) attractor. Analytical solution

Maria K. Koleva
Institute of Catalysis, Bulgarian Academy of Science
1113 Sofia, Bulgaria
e-mail: mkoleva@bas.bg; Fax: +035929712967

**ABSTRACT**

The structure of the physical and strange attractors is inherently associated with the boundedness of fluctuations. The idea behind the boundedness is that a stable long-term evolution of any natural and engineered system is possible if and only if the fluctuations that the system exerts are bounded so that the system permanently stays within its thresholds of stability. It has been established that the asymptotic structure of the physical and strange attractors is identical. Now it is found out that though the non-asymptotic behavior is universal it can be very different, namely: on coarse-graining the physical attractors can exhibit a variety of behavior while the strange attractors always have hyperuniversal properties. Yet, under certain levels of coarse-graining both physical and strange attractors matches non-asymptotically a variety of noise type behavior.

## 1. Introduction

The notion of a physical attractor is inherently consistent with the notion of the boundedness of fluctuations. The idea behind the boundedness is that a stable long-term evolution of any natural and engineered system is possible if and only if the fluctuations that the system exerts are bounded so that the system permanently stays within its thresholds of stability. Thus, the thresholds of stability match the boundaries of a finite volume in the phase space so that every phase trajectory is confined in that volume called physical attractor [1]. Further, we have proved [1] that any bounded irregular sequence (BIS) has certain asymptotic properties insensitive to the increment statistics. These properties are called $1/f^{\alpha}$-type noise. They are: (a) the power spectrum comprises a continuous band that uniformly fits the shape $1/f^{\alpha(f)}$ where $\alpha(f)$ monotonically increases starting from 1 at $f = 1/T$ ($T$ is the length of the sequence) up to $p$ as $f$ approaches infinity ($p$ is arbitrary but $p > 2$); (b) the physical (strange) attractor is non-homogeneous; (c) the Kolmogorov entropy is finite. It has been proven also that the strange attractors that are related to the simulated dynamical systems share the above properties.

The present task is to reveal the influence of the increment statistics onto the properties of the bounded irregular sequences after coarse-graining. The coarse-graining is an inevitable part of recording data at every experiment. Therefore, it is important to elucidate how it impacts the experimentally recorded time series. Taking into account that a long-term boundedness is possible if and only if the size of correlations among increments does not exceed certain finite value, it is to be anticipated that the coarse-grained structure of any BIS is universal, i.e. insensitive to the increment statistics. Indeed, an arbitrary increase of the increment correlation size certainly makes the amplitude of fluctuations to exceed the thresholds of stability. To the most surprise, it turns out that on the coarse-graining there are 3 types of non-asymptotic behavior of every BIS: (i) the low level whose major property is that the structure of any BIS matches the white noise behavior; (ii) the meso-level where the sequence of large-scaled fluctuations matches telegraphic noise behavior. At the same time, the large-scaled fluctuations retain certain specific property related to the increment statistics; (iii) macro-level where the properties of any BIS are hyperuniversal, namely insensitive to any detail of the increment statistics. It should be stressed that the hyperuniversal properties on that level of coarse-grained are due to the proof that under coarse-graining the symmetric random walk with constant step appears as the global attractor for any fractal Brownian motion. This the subject of the sec.3. On the other hand, it should be stressed that the $1/f^\alpha$ -type noise properties listed above remain unaffected by the coarse-graining. Thus, both white noise and telegraphic noise behavior are rather alias effects: they are typical for sequences whose length does not exceed certain one. The asymptotic behavior of every BIS, regardless to the level of coarse-graining, is covered always by the $1/f^\alpha$ -type noise properties established in [1].

The above separation is made on the grounds of certain Poissonian properties of the large-scaled fluctuations appearance. The particularity of that behavior is imposed by the boundedness and the lack of long-range correlations among increments. More precisely, the lack of long-range correlations among increments ensures s a uniform convergence of the average to the mean of any BIS. In turn, it provides the stationarity of the large-scaled fluctuations appearance. The existence of mean and average for any BIS is guaranteed by the Lindeberg theorem that states: each BIS has finite mean and finite variance regardless to the increment statistics [2]. The interplay between all these properties renders that each large-scaled fluctuation can be approximated by an excursion: a trajectory of a walk originating at the mean value of a given sequence at time $t$ and returning to it for the first time at time $t + \Delta$. This interval is called next duration of an excursion. The distinctive property of each excursion is that the random walk of the increment renders certain relation between the amplitude and the duration of the excursion. In sec.2 it is worked out that the boundedness provides each excursion to be loaded in a specific "embedding interval". The length of that interval is exerted as a random choice from certain range of almost equiprobable values. The presence of embedding time intervals is the warrant of the "pulse" like behavior of any BIS, namely: the BIS is a succession of excursions separated by quiescent intervals whose lengths vary with the "time" since they are exerted as a random choice from certain range. It should be stressed that this behavior is robust to the coarse-graining and it is natural on the quantum level [3]. The importance of the "pulse" like behavior is twofold: (i) it provides that the fluctuations remain bounded at any scale of averaging, respectively at any level of coarse-graining; (ii) each BIS remains a BIS on coarse-graining and thus renders the robustness to coarse-graining of the $1/f^\alpha$ -type noise properties.

## 2 ."Embedding" intervals

The target of the present and the next section are the properties of the pulse-like behavior of the coarse-grained BIS`es. We start with the excursion "embedding". Further it is proven that this property is imposed by the boundedness and is insensitive to the increment statistics. The presence of the "embedding" intervals renders that any successive excursions are separated by non-zero quiescent intervals. This is the key property for the pulse-like performing of the excursion sequences. The present task is to establish the relation between the duration of the embedding intervals and the duration of the corresponding excursions. That relation is based on the notion of an excursion: a trajectory of a walk originating at the mean value of a given sequence at time $t$ and returning to it for the first time at time $t + \Delta$. Therefore, the probability for an excursion of duration $\Delta$ is determined by the degree of correlation between any two points of a sequence. On the other hand, the probability that any two points of a sequence, separated by distance $\eta$, have the same value is given by the autocorrelation function $G(\eta)$. A generic property of the BIS`es [1] is that the autocorrelation function of any of them can be defined for sequences of arbitrary but finite length $T$. Yet, its shape is universal, namely:

$$G(\eta, T) \propto 1 - \left(\frac{\eta}{T}\right)^{\nu(\eta/T)}.$$

where $\nu\left(\frac{\eta}{T}\right)$ is an continuous everywhere monotonically decreasing between the following limits function:

$$\nu\left(\frac{\eta}{T}\right) \to p-1 \text{ as } \frac{\eta}{T} \to 0$$

$$\nu\left(\frac{\eta}{T}\right) \to 0 \quad \text{as } \frac{\eta}{T} \to 1.$$

Then, the probability that an excursion of duration $\Delta$ happens in an interval $T$ reads:

$$P(\Delta, T) = \frac{1}{T}\int_0^\Delta G(\eta, T)d\eta = P(\Delta, T) = \int_0^{\Delta/T}\left(1 - \varepsilon^{\nu(\varepsilon)}\right)d\varepsilon$$

By the use of the standard calculus [3], the integration of any power function of non-constant exponent $\alpha(x)$ ($\alpha(x)$ is an everywhere continuous function) reads:

$$J = \int_a^b x^{\pm\alpha(x)}dx = \frac{b^{\alpha(b)+1}}{1\pm\alpha(b)} - \frac{a^{\alpha(a)+1}}{1\pm\alpha(a)} \quad .$$

Thus,

$$P(\Delta, T) = \frac{\Delta}{T}\left(1 - \left(\frac{\Delta}{T}\right)^{\nu\left(\frac{\Delta}{T}\right)}\right).$$

The $P(\Delta, T)$ dependence only on the ratio $\Delta/T$ verifies the assumption that every excursion of duration $\Delta$ is "embedded" in an interval of duration $T$ so that no other excursion happens in that interval.

The next step is to work out the shape of $P(\Delta, T)$. Its role is crucial for the behavior of the excursion sequences. To elucidate this point let us consider the two extreme cases:

- $P(\Delta, T)$ is a sharp single-peaked function. It ensures a single value of the most probable ratio $\frac{\Delta}{T}$. In other words, the duration of the most probable embedding interval associated with an excursion of the most probable duration $\Delta_0$ has single value $T_0$ such that $\frac{\Delta_0}{T_0}$ is the peak value of $P(\Delta, T)$. Then, most probably the excursion sequence matches a periodic like behavior.

- $P(\Delta, T)$ has a gently sloping maximum. Then, the relation between $\Delta$ and $T$ behaves as a multi-value function: a range of nearly equiprobable values of $T$ corresponds to each most probable $\Delta$. In the course of the time this produces a random choice of the duration of the "embedding" intervals even when the sequence comprises nearly identical excursions. Then, the excursion sequence has a non-periodic, namely "pulse"-like behavior.

The establishing of the particular shape of $P(\Delta, T)$ requires the knowledge about the explicit shape of $\nu\left(\frac{\Delta}{T}\right)$. Next it is worked out on the grounds of the proof that neither BIS sustained to an arbitrary length comprises any long-ranged increment correlations. The general restriction on the increment correlation size requires an uniform contribution to the power spectrum of all scales, i.e. there are no "special" frequencies at the power spectrum.

The only factor that can modify the power spectrum is the non-constant exponent $\alpha(f)$ of its shape $1/f^{\alpha(f)}$. The boundedness requires the monotonic decay of $\alpha(f)$ in the limits $[1,p]$ not specifying its shape [1]. Our task now is to establish the shape(s) of $\alpha(f)$ that fits the lack of long-range increment correlations. In virtue of the strict monotony of the power spectrum the required criterion is that neither any its component nor any its derivative of arbitrary order has a specific contribution. Simple calculations yield that it is achieved if and only if the shape of $\alpha(f)$ is the linear decay, namely:

$$\alpha(f) = (1+\gamma f) \tag{1}$$

Eq. (1) provides that the $1/f^{\alpha(f)}$ derivative of an arbitrary order has the same sign throughout the entire frequency interval of the power spectrum. On the contrary, for any non-linear decay of $\alpha(f)$ each order derivative of $1/f^{\alpha(f)}$ changes its sign at certain frequencies. Thus, only the linear decay does not introduce any additional scale to those inherent for the increment statistics.

Because of the diffeomorfism between $\alpha(f)$ and $\nu\left(\dfrac{\Delta}{T}\right)$ it is obvious that the shape of $\nu\left(\dfrac{\Delta}{T}\right)$ reads:

$$\nu\left(\dfrac{\Delta}{T}\right) = (p-1)\left(1-\dfrac{\Delta}{T}\right). \tag{(}$$

The plot of $P(\Delta,T)$ with the above shape of $\nu\left(\dfrac{\Delta}{T}\right)$ shows that it has a gently sloping maximum: indeed, the values of $P(\Delta,T)$ in the range $\dfrac{\Delta}{T}\in[0.25,0.4]$ vary by less than 7%. Outside this range $P(\Delta,T)$ decays sharply: Fig1. Thus, though $P(\Delta,T)$ is a single-valued function, it provides a multi-valued relation between the most probable values of $\Delta$ and $T$, namely: a certain range of nearly equiprobable values of $T$ is associated with each $\Delta$. In the course of the time the multi-valued relation is exerted as a random choice of the duration of the "embedding" intervals even when the sequence comprises identical excursions. Thus, the multi-valued relation between $\Delta$ and $T$ provides that any coarse-grained BIS is again a BIS. Then, the $1/f^{\alpha}$-type noise properties are robust the coarse-graining.

## 3. Symmetric random walk as global attractor for the fractal Brownian motion

Every BIS comprises fine-structure fluctuations superimposed on the large-scaled ones. The former are created by the short-range increment walk and the latter appear as successive excursions from the mean value of the BIS. It is to be expected that the fine structure fluctuations strongly depend on the particularities of the increment statistics while the large scaled ones have certain generic properties robust to that statistics. One such property is that there is certain relation between the amplitude of each fluctuation and its duration. Next it is worked out for the case when the increment size is much smaller than the thresholds of stability. Then the fluctuations can be approximated by excursions created by continuous fractal Brownian motion of the increment walk in the course of the time. The latter renders certain relation between the amplitude $A$ and the duration $\Delta$ of the excursion, namely: $\sqrt{\langle A^2 \rangle} \propto \Delta^{\beta(\Delta)}$, $\beta$ is set by the particularity of the increment statistics; the averaging is over the sample realisations. The dependence of $\beta$ on $\Delta$ comes from the interplay between the finite radius of the increment correlations $a$ and the amplitude of the excursion itself that is limited only by the thresholds of stability. Next it is proven that each time when $a << A$ the fractal Brownian motion uniformly approaches the symmetric random walk with a constant step. Indeed, since the increment size is permanently bounded, the mean square deviations (m.s.d.) of all the trajectories of the same number of steps $N$ is confined in a finite range. Therefore, these m.s.d. also form a BIS. Further, according to the Lindeberg theorem [2] the latter has finite mean and finite variance and thus it is a subject to the Central Limit Theorem. As a result, the m.s.d. are Gaussianly distributed. In turn, the sizes of the successive trajectories of a given $N$ are predominantly equal to the corresponding mean. So, the increment correlations are to be associated with a single trajectory whose number of steps $m$ is related to the increment correlation size and the exponent $\tilde{\beta}$ is specific to the increment statistics. In other words, the increment correlations creat "blobs" whose size is $\sqrt{\langle a^2 \rangle} \propto m^{\tilde{\beta}}$. Then, the large excursions can be approximated by a symmetric random walk with constant step equal to the size of the blobs. Thus the dependence of any large scale excursions on its duration reads:

$$\sqrt{\langle A^2 \rangle} \propto N^{0.5} m^{\tilde{\beta}}$$

where $N$ is the number of the blobs.
It is obvious that when $N >> m$ the dependence tends to:

$$\sqrt{\langle A^2 \rangle} \propto N^{0.5} a$$

where $a$ is considered constant independent of $N$. So, the symmetric random walk with constant step appears as the global attractor for any fractal Brownian motion.

It is worth noting that the finite size of the "blobs" ensures the uniform convergence of the average to the mean of the original BIS. Indeed, the distinctive property of any fractal Brownian motion is that any power $\beta \neq 0.5$ arises from an arbitrary correlation between the current increment $\mu_i$ and the corresponding step $\tau_i$. Then the average $A$ reads:

$$A = \sum_{i=1}^{N} \mu_i(\tau_i) \tau_i = \sum_{i=1}^{N} (-1)^{\gamma_i} \tau_i^{\beta_i} \qquad (2)$$

and correspondingly the m.s.d. :

$$\langle A^2 \rangle \propto \left\langle \sum_{i=1}^{N} (\mu_i(\tau_i)\tau_i)^2 \right\rangle = \left\langle \sum_{i=1}^{N} \tau_i^{2\beta_i} \right\rangle$$

where the averaging is over the different samples of the trajectory; The property of the above relations is that whenever the probabilities for $\gamma_i$ odd and even are not permanently equal there is a correlation between the increment and the corresponding step. So $A$ is certainly non-zero which immediately makes that the deviation from the mean non-zero. Moreover, eq.(2) yields that $A$ can become arbitrarily large on increasing $N$. On the contrary, a permanent equal probability for $\gamma_i$ odd and even means independence from one another of the increments and the steps. It yields $A = 0$ which guarantees the uniform convergence of the average to the mean.

## 4. Statistics of the coarse-grained BIS`es

Our first task is to elucidate that the excursion appearance meets the three generic features of a Poissonian process, namely: (I) the excursion appearance is a stationary process. (ii) the successive excursions are independent from one another events; (iii) no more than one event can be developed at any instant. The stationarity of the excursion sequences is guaranteed by the following interplay between the boundedness and the lack of any long-ranged increment statistics. Indeed, the Lindeberg theorem [2] states that any BIS has finite mean and finite variance insensitively to the increment statistics. Obviously, the coarse-graining affects neither the presence nor the value of the mean. Further, the lack of long-ranged increment correlations guarantees the uniform convergence of the average to the mean. This, in turn verifies the stationarity of the excursion appearance. This along with the embedding of each excursion in a larger interval is the warrant that the successive excursions are independent from one another events.

On the other hand, both the boundedness and the lack of long-ranged increment correlations render that every BIS is a subject to the central limit theorem. So, the departure from the mean obeys the Gaussian distribution. Our task now is to elucidate the interplay between the Poissonian properties and the Gaussian ones under the coarse-graining. It is obvious that the coarse-graining makes all the excursions of the amplitude less then the size of coarse-graining $A_{cgr}$ to contribute to the quiescent intervals. Then, the excursion of a given amplitude appears in the time course with the probability that reads:

$$P(A) = \frac{A^{1/\beta(A)} \exp(-A^2/A_\sigma^2)}{(1+\eta)A^{1/\beta(A)} \exp(-A^2/A_\sigma^2) + \int_0^{A_{cgr}} A^{1/\beta(A)} \exp(-A^2/A_\sigma^2)dA} \qquad (11)$$

The required probability $P(A)$ is given by the ratio between the weighted duration $A^{1/\beta(A)} \exp(-A^2/A_\sigma^2)$ of an excursion of amplitude $A$ and the weigthed length of the corresponding embedding interval. The latter comprises the weighted duration of the excursion itself, the weighted duration of the natural embedding interval expressed through $\eta$ and the weighted duration of the quiescent interval due to coarse-grainning $\int_0^{A_{cgr}} A^{1/\beta(A)} \exp(-A^2/A_\sigma^2)dA$; $A_\sigma$ is the variance of the BIS and in the present consideration it is a parameter. The Poissonian properties of the excursion appearance ensure that $P(A)$ has the same value at every point of the sequence. It should be stressed that eq.(11) holds only for $A \geq A_{cgr}$. Otherwise $P(A) \equiv 0$ since all the excursions of amplitude less than the size of coarse-graining $A_{cgr}$ are "smoothed out" and contribute to the quiescent intervals.

$P(A)$ exhibits 3 different types of behavior with respect to the ratio $A/A_\sigma$. In the case $A << A_\sigma$ the following approximation holds:

$$J(A) = \int_0^{A_{cgr}} A^{1/\beta(A)} \exp(-A^2/A_\sigma^2)dA \propto A_{cgr}^{1+1/\beta(A_{cgr})} \quad .$$

For the case when $A_{cgr} \leq A << A_\sigma$ it yields:

$$P(A) \approx \frac{1}{1+\frac{A}{A_\sigma}} \propto 1.$$

So, in this case the fluctuations of different amplitude are equiprobable which is a distinctive property of the white noise behavior. Therefore, a BIS behaves as a white noise sequence. Yet, a warning is necessary: (I) this behavior is available at up to certain finite length of the sequence. That length is proportional to the inverse of the probability for appearance of an excursion of amplitude $A$. So, it is of the order of $\exp\left(\dfrac{A^2}{A_\sigma^2}\right)$. The asymptotic behavior of the sequence is covered by the $1/f^\alpha$ - type noise properties listed in the Introduction.

The other extreme case is $A_{cgr} \gg A_\sigma$. The approximation of $J(A)$ in this case reads:

$$J(A_{cgr}) \approx const.$$

Then for $A_{cgr} \leq A \gg A_\sigma$  $P(A)$ becomes:

$$P(A) \propto A^{1/\beta(A)} \exp\left(-\dfrac{A^2}{A_\sigma^2}\right). \tag{4}$$

In this case the excursion sequence behavior matches telegraphic noise, namely the excursions are almost identical separated by large quiescent intervals. An immediate result of eq.(4) is that the amplitude of these excursions is set by the most probable one, namely $A = A_{cgr}$. Thus, the length of the sequence that matches the telegraphic noise behavior is proportional to $\exp\left(\dfrac{A_{cgr}^2}{A_\sigma^2}\right)$. The asymptotic behavior of the sequence is again covered by the $1/f^\alpha$ - type noise properties listed in the Introduction. It is worth noting that this type of telegraphic noise differs from the dichotomic one in its power spectrum. The dichotomic noise has an exponential power spectrum while the telegraphic noise that comes from a BIS has the power spectrum of the shape $1/f^{\alpha(f)}$. The exponent in the former case comes from the second kind discontinuities while only first kind discontinuities are available for the bounded sequences.

The behavior of any coarse-grained BIS is inherently related to the incremental statistics trough the explicit dependence of $P(A)$ on $\beta(A)$ in eq.(3). However, when the size of coarse-graining is much larger than the size of the symmetric random walk $\beta(A)$ turns constant equal to $0.5$. Then, $P(A)$ gets insensitive to the details of the increment statistics and all the characteristics of the coarse-grained BIS become hyperuniversal, i.e. insensitive to any particularity of the increment statistics. The hyperuniversal behavior of any BIS appears whenever $a \ll A_{cgr}$. Since $a$ is finite, there is always large enough size of coarse-graining such that $a \ll A_{cgr}$. So, at large enough sizes of coarse-graining the behavior of the BIS is to be hyperuniversal. It is worth noting, however, that since $a$ and $A_\sigma$ are independent parameters, it is possible that $a \ll A_\sigma$. Then the behavior of the corresponding BIS is hyperuniversal at almost every level of coarse-graining.

These considerations yield an apparent distinction between the physical attractors and the strange attractors that originates from the simulated dynamical systems. The strange attractors arises at unstable solutions where the inevitable round-off at simulation is amplified due to positive Lyapunov coefficient [1]. Since the round-off to the higher and the lower value is permanently equiprobable, the power $\beta$ at the fractal Brownian motion is $0.5$ for the excursions of all sizes. Hence, the strange attractors have hyperuniversal structure at every level of coarse-graining. On the contrary, the physical attractors exhibit variety of behavior under the coarse-graining.

**Conclusions**

The major physical result of the present work is that the boundedness is incompatible with long-range increment correlations. This result gives rise to several very important consequences:

(I) $1/f^{\alpha}$-type noise properties of each BIS are scale-free and robust to coarse-graining;

(ii) pulse-like behavior is also typical for every BIS and is robust to coarse-graining. Being natural on the quantum level [3] it is self-reproduced on every level of coarse-graining through the existence of embedding intervals introduced by the boundedness and the coarse-graining. The major feature of the embedding introduced by the boundedness is that its length is a random choice from certain range at any point of the sequence. In turn, the variability of that length ensures the irregularity of every BIS even of that comprising identical excursions. This lays a base to the insensitivity of the $1/f^{\alpha}$-type noise properties to the coarse-graining;

(iii) the symmetric random walk with constant step appears as the global attractor for any fractal Brownian motion. In turn, the size of that step provides the scale beyond which both the structure and the succession of the excursions are insensitive to the incremental statistics. Below that scale these properties are affected by the incremental statistics through the power exponent β of the fractal Brownian motion that approximates the excursions.